\documentclass[12pt]{article}
\usepackage{doublespace}
\usepackage{multicol}
\usepackage{amsmath}
\usepackage{array}
\def\@biblabel#1{[#1]\hfill}
\begin{document}           

\begin{spacing}{1.3}


\newcommand{\npi}{\hspace{-0.5cm}} 
\newcommand{\phy}{Department of Physics}
\newcommand{\ssi}{Solid State Institute}
\newcommand{\technion}                                         
{Technion--Israel Institute of Technology, Haifa 32000, Israel}
\newcommand{\smq}{\mbox{$\simeq$}}
\newcommand{\pr}[1]{Phys. Rev. {\bf {#1}}}        
\newcommand{\prb}[1]{Phys. Rev. {\bf B {#1}}}     
\newcommand{\prl}[1]{Phys. Rev. Lett. {\bf {#1}}} 
\newcommand{\fg}[1]{Fig.~\ref{#1}}

\begin{center}
\npi {\large {\bf
Radiative Lifetimes of Single Excitons in Semiconductor 
Quantum Dots- Manifestation of the Spatial Coherence Effect
}}\\ 

\vspace{3mm}

{\bf E. Dekel$^a$, D.V. Regelman$^a$, D. Gershoni$^a$, E. Ehrenfreund$^a$},
{\bf W.V. Schoenfeld$^b$ and P.M. Petroff~$^b$ }\\
$^a$\phy\ 
\technion \\
$^b$Materials Department, University of California, Santa Barbara, CA 93106, 
USA 

\end{center}


\vspace{2mm}

Using time correlated single photon counting
combined with temperature dependent diffraction limited confocal 
photoluminescence spectroscopy we accurately determine, for the first time, 
the intrinsic radiative lifetime of single excitons confined within
semiconductor quantum dots. 
Their lifetime is one (two) orders of magnitude longer than the intrinsic
radiative lifetime of single excitons confined in semiconductor 
quantum wires (wells) of comparable confining dimensions. 
We quantitatively explain this long radiative time in 
terms of the reduced spatial coherence between the confined 
exciton dipole moment and the radiation electromagnetic field.

\vspace{5mm}


\newpage

\setlength{\columnsep}{6mm}


In intrinsic direct bandgap semiconductors the interaction of 
excitons with the electromagnetic
radiation field leads to a stationary state called exciton-polariton.
This elementary excitation is a coherent state of the entire 
macroscopic three dimensional (3D) crystal, and as such it does not 
decay radiatively \cite{Hopfield}.
In geometrically restricted semiconductor quantum structures such as quantum 
wells (QWs), wires (QWRs) or dots (QDs) the situation is different. 
In these systems, the translational symmetry of the crystalline potential 
is removed in one, two or three directions, respectively.
As a result, the three dimensional electromagnetic radiation field interacts 
with excitons of lower dimensionality (2D, 1D or 0D, respectively).
This dimensionality difference gives rise to exciton-polariton modes 
that are not stationary states. These modes have finite 
energy width, $\Delta E$, and consequently they do decay radiatively,
[2-5] 
with intrinsic radiative lifetime $\hbar/\Delta E$, 
which, for typical $GaAs/AlGaAs$ QWs, 
amounts to a few tens of picoseconds\cite{Andreani-QW,Devaud}, while for
QWRs of comparable dimensions it is considerably 
longer.\cite{Citrin-prl,Akiyama,mkprb}.
This counter-intuitive result was explained in terms of the reduced 
exciton-coherence volume, imposed by the additional lateral 
confinement.\cite{Agranovitch,Citrin-prl}

The arguments of this exciton-polariton theory explain also the considerably longer 
photoluminescence (PL) decay times which were measured in these
systems.[7-9]
They result from the thermal distribution of the excitonic 
population, where only a small fraction occupies the optically active 
part of the available excitonic phase space.\cite{Andreani-QW,Citrin-prl}
Consequently, the dimensional dependent density of energy states reveals itself
in the temperature dependence of the radiative decay times. Indeed, 
both theory and experiment show that these are proportional to the temperature, 
$T$, in QWs \cite{Andreani-QW,Feldman} and to $\sqrt{T}$ in 
QWRs.\cite{Citrin-prl,Akiyama,mkprb}
When these considerations are applied to QDs, longer intrinsic radiative 
lifetimes are expected,\cite{Citrin-QD} and 
their characteristic $\delta$-function like density of energy states leads 
to $T$ independent radiative decay times.
Though numerous experimental studies of the PL decay times from QD
assemblies \cite{Times-array} and more recently from single QDs \cite{Times-single}, 
have already been carried out, this theoretically expected behavior has 
never been observed.

In this Letter we report on the first experimental determination of 
the radiative lifetime of single excitons confined within semiconductor QDs.
We determine this fundamental property of 0D excitons by applying time 
correlated single photon counting combined with confocal microscopy 
to single self assembled quantum dots (SAQDs). The temporal evolution of the 
PL emission following pulse excitation at various intensities is measured 
and the PL intensity is quantitatively analyzed taking into account the 
interaction between a number of excitons (multiexcitons) 
confined within one QD.\cite{erez-prl,erez-prb1} 
Our intensity and temporal analysises unambiguously determine the intrinsic 
radiative lifetime of single QD excitons. 
We show that they are few nanoseconds (ns) long and $T$ independent.

We studied two types of SAQDs. Both were fabricated by 
molecular beam epitaxial deposition of a defect free coherently strained 
epitaxial layer of $InAs$ on either directly on a $GaAs$ layer by the 
partially covered islands method\cite{Garcia-APL} (PCI sample) or on an 
$AlGaAs$ host layer
(AHL sample).\cite{erez-prl} 
The layer sequences, compositions and widths 
are reported elsewhere.\cite{Garcia-APL,erez-prl}
During the growth of the strained layers, the samples were not rotated. 
Thus, a gradient in the QDs density was formed and low density areas, 
in which the average distance between neighboring QDs is
larger than the spatial resolution of our 
microscope were found on their surfaces.
A key difference between the two samples is their non-radiative decay rates.
The presence of $Al$ within the SAQDs' host layers of the AHL sample gives rise
to very efficient non-radiative recombination centers,\cite{ITAY} close to 
the SAQDs. Therefore, at low excitation densities their PL decay time is 
extremely short.\cite{erez-prb1} 
It is shown below that in spite of this difference, the determined radiative 
lifetimes of single excitons in both samples are roughly the same, 
as expected from the dimensional considerations.

The PL was excited by a synchronously mode-locked 3 ps pulsed 
dye-laser through a low temperature confocal microscope, which is described 
elsewhere.\cite{erez-prb1} The system provides spectral and spatial resolutions
of 0.2 meV and \smq0.5 $\mu m$, respectively.
In addition, a thermoelectrically cooled avalanche silicon photodiode was used
for photon counting with temporal resolution of 200 ps.

In Fig.1a(1b) we present by solid lines temporally integrated PL 
spectra from a single PCI (AHL) SAQD for various $\ge 1.75$ eV pulsed 
excitation powers. At the lowest power PL spectrum of Fig.1a a single 
narrow spectral line is observed at energy of 1.284 eV. 
We denote this line by $A_1$. Its linewidth is limited by our setup resolution.
As the excitation power increases, 
a satellite spectral line ($A_2$) appears 3.5 meV below $A_1$
and a higher energy spectral line, ($B_1$) emerges at 1.324 eV, 40 meV
above $A_1$. At yet higher excitation power line $B_1$ develops a 1.5 meV 
lower in energy satellite ($B_2$). Further increase in the excitation power 
results in an increase in the number of satellites which gradually form 
spectral emission bands to the lower energy side of the lines $A_1$ and $B_1$. 
In addition, new higher energy groups of lines,
$C$, $D$ and $E$, respectively, gradually emerge and develop their 
own satellites and lower energy bands.
The various lines are marked by letters which represent
the energy group to which they belong and numerical subscripts which
represent their appearance order with increasing the excitation power. 
As the power increases, after reaching maximum, the intensity of the 
emission from the various spectral lines 
ceases to increase and it remains constant with further increase in the power.
This is in marked difference from their behavior under continuous 
wave (cw) excitation mode, where they gradually disappear.\cite{erez-prb1} 

The behavior of the PL spectra from the AHL sample (Fig.1b) is similar, 
except for two major differences: First, only two groups of lines, 
A and B, are observed. Second, the satellite spectral line $A_2$ appears 
before the line $A_1$. 
We explain the first difference by the different dimensions of the two dots. 
Whereas the first, larger one, has 4-5 confined electron levels,
the second one has only two. The second difference is due to the efficient
non-radiative decay process associated with the $Al$ atoms in the AHL sample.
At low excitation densities, before it is saturated,\cite{erez-prb1} 
it inhibits the observation of the line $A_1$. 
Our model simulations\cite{erez-prb1} are presented in the figures by 
gray lines.

In Fig.2a(2c) we present the PL intensity of the single SAQD spectral 
line $A_1$($B_2$) from the PCI(AHL) sample as a function of time after the 
excitation pulse. 
The PL transients were measured at 20K for various excitation powers.
In general, for each spectral line there are two distinct time domains.
In the first, the emission intensity rises and in the second, it decays.
As a rule we noted (not shown) that the lower is the energy of a
given spectral group of lines, the longer are their rise and decay times.
Within a given group of lines, however, lower energy lines rise and 
decay faster. Our model calculations\cite{erez-prb1} for the $A_1$($B_2$)
line of the PCI(AHL) sample 
are presented for comparison in Fig.2b(2d).
Clearly, the rise time of a particular spectral line, strongly depends on 
the excitation power. The higher the power is, the longer is the rise time, 
while the decay time is hardly affected. 
We note that the temporal behavior of the spectral lines from the different 
samples is very similar. The main difference is the density of excitation 
needed to observe the PL from the second sample. 
This difference, like the difference in the order by which the 
spectral lines appear with increasing excitation power 
is explained by the vast efficiency difference between the 
non-radiative recombinations in the two samples.
In the inset to Figs.2a(2c) we display by symbols the 
measured rise time of the spectral line $A_1$($B_2$), as a function of the 
excitation power. Our model calculations are presented for comparison in 
the inset to Fig.2b(2d). Similar measurements and calculations 
are presented in Fig.3a and Fig.3b, respectively, for the spectral 
line $A_1$ of an SAQD from the PCI sample, at higher sample temperature of 70K. 

The PL spectrum of a semiconductor QD excited by a short laser pulse 
is due to optical transitions between multiexciton states.\cite{erez-prb1}
The excitation generates population of electron-hole (e-h) pairs within the 
dot. The many e-h pairs form a correlated quantum state which we define 
as a multiexciton. The photoexcited carriers reach thermal distribution on a 
very short time scale compared with the time required for radiative 
recombination of an e-h pair. 
Thus, at cryogenic temperatures, the specific excited multiexciton rapidly relaxes to its 
ground level. A radiative annihilation of one of the e-h pairs that compose the 
multiexciton may then occur by the emission of a photon, which carries the energy 
difference between the annihilated multiexcitonic ground level and that of the newly 
generated one. 
The new multiexciton, with one less e-h pair
quickly relaxes to its ground level and eventually recombines by losing yet
another e-h pair. The process continues sequentially, until the
dot remains empty of excitons.
For a small number of QD excitons ($\le 2$), the probability of non-radiative
recombination becomes increasingly important. First, since the non-radiative 
decay channels are no longer screened by the carriers in the QD and 
its vicinity. Second, since the radiative recombination has fewer channels 
and its rate is much slower.\cite{erez-prb1}

In order to understand the spectral evolution of the SAQD PL
with excitation power and with time, a knowledge of the various multiexcitons 
energies and recombination rates is required. Our model provides a 
straightforward way for calculating these energies using a minimal set of 
parameters (i.e., the single carrier energy levels and the exchange energies 
between these levels), which can be directly obtained from the PL spectrum 
itself.\cite{erez-prb1}.
The recombination rates are given by the model in terms of the QD 
single exciton radiative lifetime ($\tau_1$) through the use of the 
dipole approximation, together with symmetry and total angular momentum 
conservation considerations. These energies and rates are used for 
calculating the spectra in Fig.1, where a Gaussian spectral 
broadening of 0.5 meV for each discrete multiexcitonic level was used, 
and the statistical nature of the pulse photogeneration was taken into account.
Thus, $<N_X>$ in Fig.1 represents the average number of
photogenerated pairs (or the multiexciton order) in the QD by each laser pulse.
Since the spectra in Fig.1 are temporally integrated, each multiexciton
of order $\le N_X$ contributes exactly one radiative 
recombination to the measured spectrum.
From this simple argument, it follows that the comparison between the calculated
spectra of Fig.1 and the measured spectral density of counts per pulse at 
saturation, directly yields the collection efficiency of our setup.
This calculated number is almost model independent since the number of
photons per recombination is easily obtained from total spin conservation 
considerations.\cite{erez-prl,erez-prb1,zunger-condmat}
The efficiency obtained from Fig.1, \smq5$\times 10^{-5}$, does not depend on
the efficiency of excitation or on the non-radiative processes and it 
is in a very good agreement with our estimates based on measuring the reflected laser light.

We use the multiexcitons calculated recombination rates as input parameters
to a set of coupled rate equations which describe the temporal behavior of the pulsed 
photoexcited single QD. The rate equations are analytically solved
for the cw and pulsed excitation modes.\cite{erez-prb1} 
The cw solution yields a direct determination of the radiative 
lifetime 
$\tau_1$. For this task one uses the determined efficiency of the 
experimental system and the cw excitation power dependence of the 
PL spectrum. As mentioned above, when the cw 
excitation power increases the various SAQD PL discrete spectral lines 
reach maximum emission and then lose intensity
with further increase in the power
(not shown).\cite{erez-prl,erez-prb1} 
The steady-state number of QD excitons when a certain line reaches 
maximum is thereby given by comparison with the analytic solution 
of the rate equations. The calibrated emission intensity at 
these conditions, accurately determines $\tau_1$.\cite{erez-prb1}

Another independent way of determining $\tau_1$ is demonstrated by the
curves in Figs.2b and 2d which represent the solution to the rate
equations of the pulsed excitation mode.
Two parameters are used for generating all the simulated transient 
curves. Namely, 
$\tau_1$ and the ratio between it and the non radiative 
lifetime ($\tau_{nr}$). 
From the comparison between the measured data 
(Figs.2a,2c and their insets) and the
calculated ones (Figs.2b, 2d and their insets) we find that the best 
fit for the PCI sample is obtained when $\tau_1/\tau_{nr} \smq 5$.
For this and higher ratios, $\tau_{nr}$ can be directly obtained 
from the PL decay time of $A_1$, which exponentially decays within
$0.8\pm0.1$ ns ($\tau$ in Fig.2a). 
Consequently, $\tau_1$ that we deduced for the PCI SAQDs at 
this temperature is $4\pm 1$ ns, in very good agreement with its 
PL efficiency determination.\cite{erez-prb1}.

We state here that the determination of $\tau_1$ is not very 
sensitive to $\tau_{nr}$ since it is determined mainly from the 
power dependence of the rise time. 
Thus, variations in $\tau_1/\tau_{nr}$ ratio from zero to almost infinity
yields only up to a factor of two change in the deduced $\tau_1$.
Similarly, we determined $\tau_1$ of the AHL SAQDs to be $5\pm 1$ ns, 
and more than 2 orders of magnitude longer than their $\tau_{nr}$. 

The decay time of the PL in the PCI sample is found to rapidly 
increase with $T$, while the rise time hardly changes.
This is evident from the PL decay time of the $A_1$ line at 
70K ($\tau$ in Fig.3a), which is $5\pm 1$ ns long. 
We believe that this is due to reduction in the non-radiative rate 
due to thermal activation of trapped charges that act as non 
radiative centers in the vicinity of the SAQD.
The best fits in Fig.3b were obtained with $\tau_1/\tau_{nr}\ll 1$.
This means that at this temperature the non-radiative rate can be 
neglected, and the radiative time which determines the PL rise time, now 
determines its decay time as well.
The AHL sample non-radiative rates are activated with increasing $T$ and they 
become even faster\cite{ITAY}. They render the temporal resolution of the 
emission from discrete PL lines at elevated $T$, nearly impossible.

The lifetime that we obtain for both samples, using these two 
independent experimental methods, amounts to 4-6 ns, 
and it is temperature independent. This is much longer than the 
measured PL decay times from single 
SAQDs\cite{Times-single} and from SAQD assemblies\cite{Times-array}, 
where typically, like here, low temperature sample dependent 
decay times of subnanosecond are reported. 
We show here that this is because at low excitation densities the 
decay time depends crucially on the non-radiative recombinations, 
while at high excitation densities one measures faster decays of 
multiexcitons. Our approach, which is independent of the measured 
PL decay time, clearly circumvent these impediments. 
We demonstrated that only at 70K, when the non-radiative
decay process of the PCI sample is blocked, the PL decay is purely 
radiative and $\tau_1$ can be directly obtained from its measurement.
We also demonstrated that even for 
the QDs in the AHL sample 
long radiative lifetimes are deduced, in spite of their very fast
non radiative decay rates. 

The QD single exciton radiative lifetime is much longer than the 
intrinsic radiative time of higher dimensional systems such as 
wires\cite{Citrin-prl,mkprb,Akiyama} and wells.\cite{Andreani-QW,Devaud,Feldman} 
Following the arguments of Andreani\cite{Andreani-QW}, and Citrin\cite {Citrin-prl} 
who calculated the intrinsic radiative lifetimes of single excitons in semiconductor QWs
and QWRs, respectively, we find for 
excitons in spherical QDs: 
%
\begin{equation}
\tau_{0}^{QD}= 
{n \, m_{0} \, c }
/
{ \ e^2 \,f \, k_{ex}^2 }
~,\label{1}
\end{equation}
where $n$ ,$m_0$, $e$ and $c$ are the material index of refraction, electronic mass, electron
charge and speed of light, respectively, $f$ is the confined exciton
oscillator strength and $k_{ex}=n\omega/c$ is the exciton wavevector in the material.
For exciton energy $\hbar \omega$=1.35 eV and $f$=0.5 for QDs of these 
dimensions\cite{Takagahara}, we get $\tau_{0}^{QD}$=4.2 ns,
in reasonable agreement with our temperature independent experimental findings. 
We therefore conclude that the long QDs single exciton radiative
lifetimes that we measured are due to the small spatial phase coherence 
between the 3D confined excitonic dipole and the electromagnetic radiation field in the 
semiconductor matter.

\hspace{0.5cm} {\it Acknowledgments} -- 
The work was supported by the Israel Science Foundation 
and by the Israel-US Binational Science Foundation.


\npi {\bf Figure Captions}

\newcounter{fg}
\refstepcounter{fg}
\label{fig1}
\refstepcounter{fg}
\label{fig2}
\refstepcounter{fg}
\label{fig3}
     
\begin{list}
{\fg{\arabic}}{\usecounter{fg}}

\item[\fg{fig1}]
a)(b))PL spectra from a single SAQD from the PCI(AHL) sample
for various pulsed excitation powers. The gray lines and 
right axes represent our model simulations.

\item[\fg{fig2}]
The intensity of line $A_1$($B_2$) from a single SAQD of the PCI(AHL) sample 
$vs.$ time after the excitation pulse measured a)(c)) for various excitation 
powers and calculated b)(d)) for various average number of photogenerated 
QD excitons, $<N_x>$. The insets to a) and c) ( b) and d)) display the 
measured (calculated) rise times of the PL, $vs.$ excitation power( $<N_x >$).

\item[\fg{fig3}]
The intensity of line $A_1$ from SAQD from the PCI sample $vs.$ time at 70K. 
a)(b))Measured (calculated) for various powers ($<N_x >$ values).
The inset display the measured (calculated) rise time of the PL
$vs.$ the excitation power ($<N_x>$). 
\end{list}


\end{spacing}

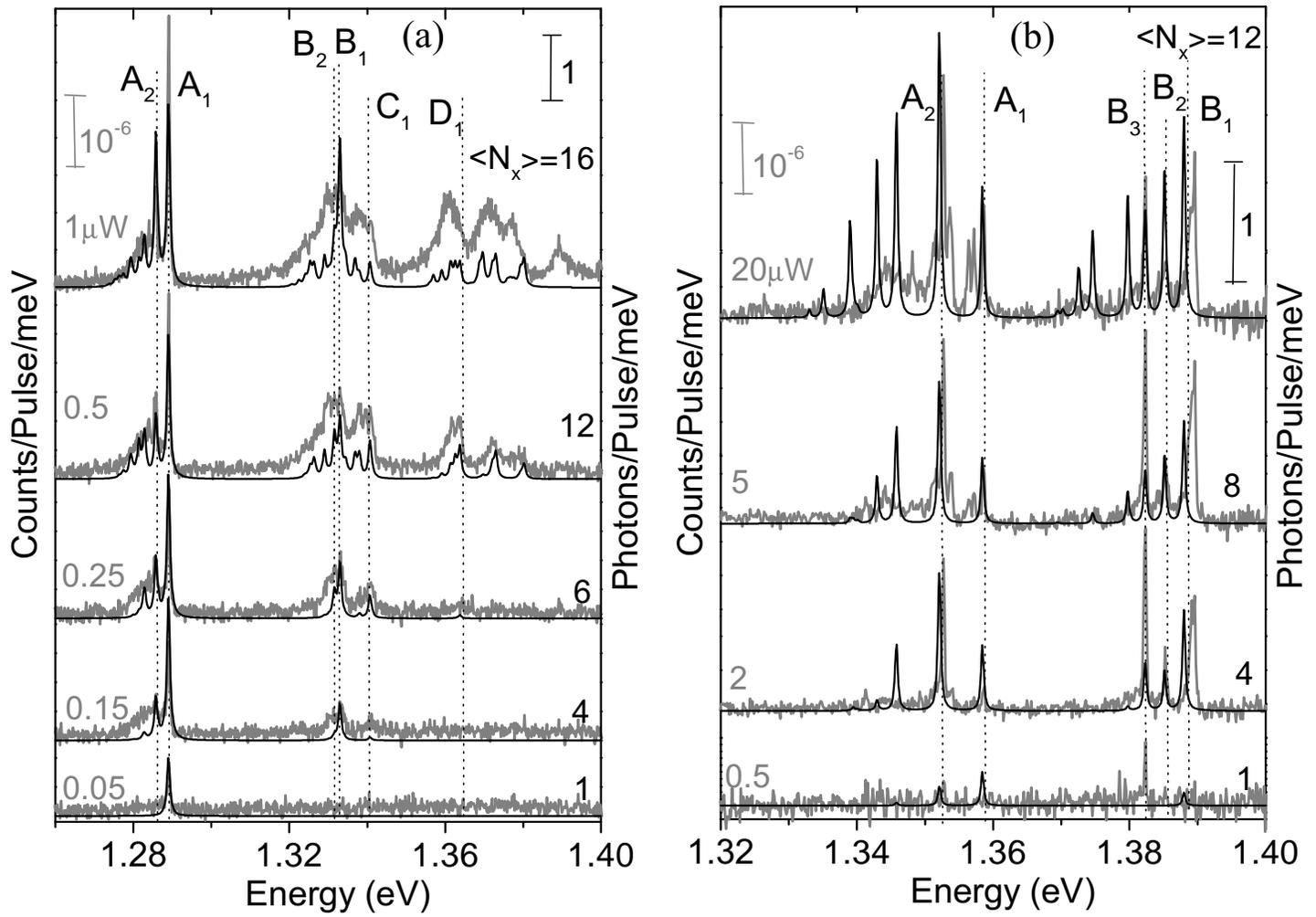
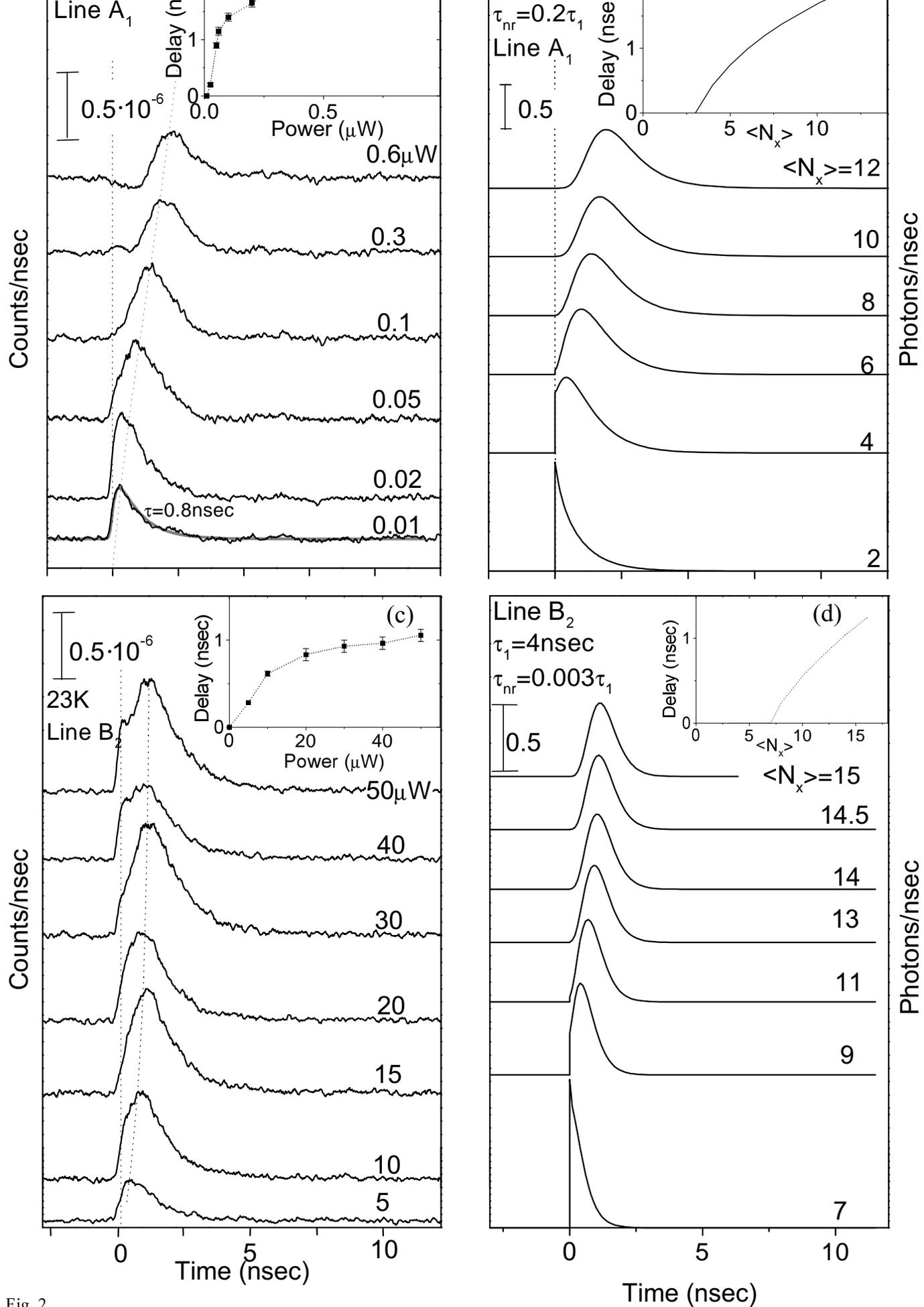
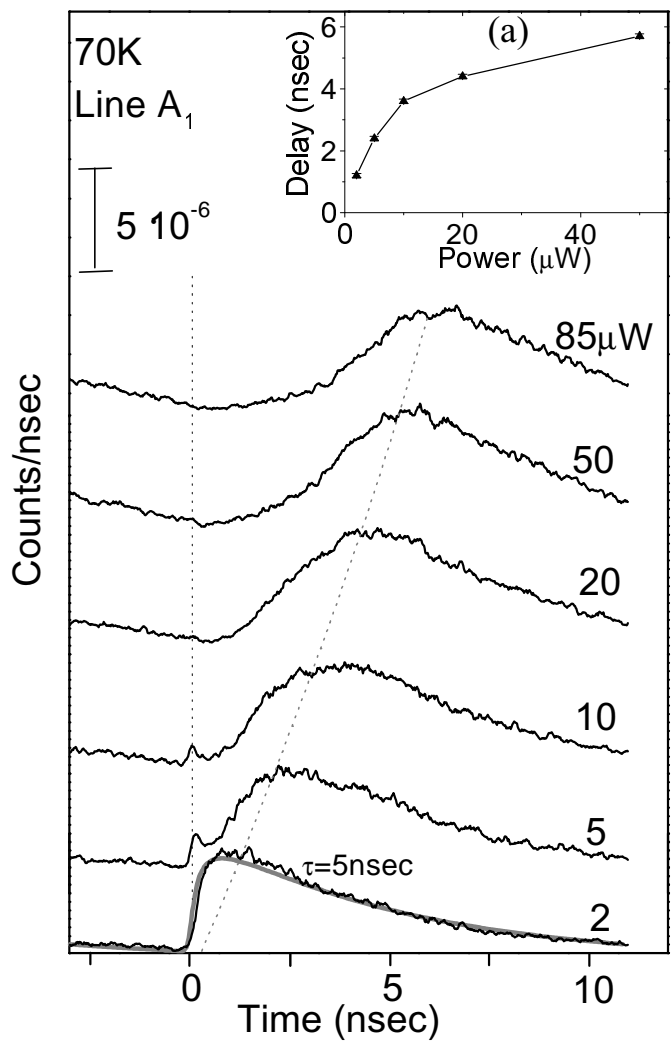
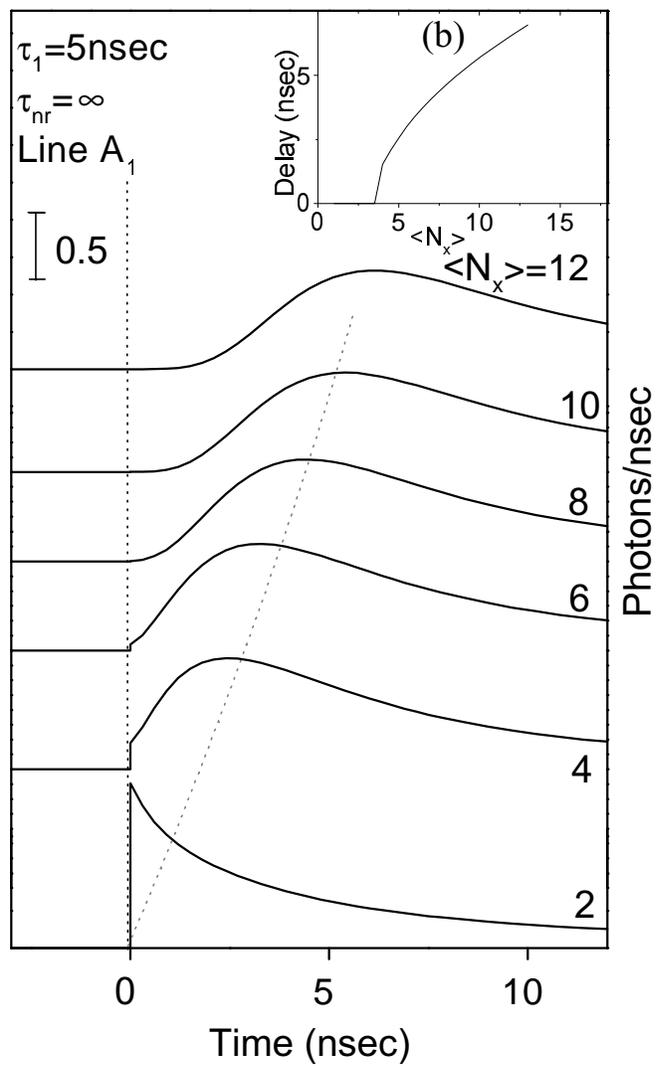
\end{document}